\documentclass[10pt,conference]{IEEEtran}
\IEEEoverridecommandlockouts
\usepackage{amsmath,amssymb,amsfonts}
\usepackage{algorithmic}
\usepackage{graphicx}
\usepackage{textcomp}
\usepackage{xcolor}
\usepackage{minted}
\usepackage{booktabs}
\usepackage{multirow}
\usepackage{graphicx}
\usepackage{subcaption}
\usepackage{tabularx}
\usepackage[table, dvipsnames]{xcolor}
\usepackage{xurl}

\usepackage{listings}
\def\BibTeX{{\rm B\kern-.05em{\sc i\kern-.025em b}\kern-.08em
    T\kern-.1667em\lower.7ex\hbox{E}\kern-.125emX}}
\begin{document}

\title{NL in the Middle: Code Translation with LLMs \\ and Intermediate Representations}

\author{\IEEEauthorblockN{Chi-en Amy Tai\IEEEauthorrefmark{1},
Pengyu Nie, Lukasz Golab, and
Alexander Wong}
\IEEEauthorblockA{University of Waterloo, Canada \\
Email: \IEEEauthorrefmark{1}amy.tai@uwaterloo.ca}}

\maketitle

\begin{abstract}
Studies show that large language models (LLMs) produce buggy code translations. One promising avenue to improve translation accuracy is through intermediate representations, which provide structured guidance for the translation process.
We investigate whether LLM-based code translation can benefit from intermediate representations, specifically in the form of natural language (NL) summaries and abstract syntax trees (ASTs).
Since prompt engineering greatly affects LLM performance, we consider several ways to integrate these representations, from one-shot to chain-of-thought (CoT) prompting. 
Using Open GPT4 8X7B and specialized StarCoder and CodeGen models on popular code translation benchmarks (CodeNet and AVATAR), we find that CoT with an intermediate NL summary performs best, with an increase of 13.8\% and 6.7\%, respectively, in successful translations for the best-performing model (Open GPT4 8X7B) compared to the zero-shot prompt.
\end{abstract}


\section{Introduction}
Code translation is the task of converting code from one programming language (e.g., Java) to another (e.g., Python)~\cite{weisz2022better}. This task is important in software as organizations often need to migrate legacy systems to newer technologies, improve code interoperability, or leverage language-specific features and libraries~\cite{pan2024lost}. However, manual code rewriting is a labor-intensive process that is inherently prone to human error, thereby impacting both efficiency and accuracy. While early statistical efforts~\cite{oda2015learning} have recently given way to methods based on large language models (LLMs), studies show that LLMs still produce buggy translations when using a zero-shot prompt~\cite{pan2024lost}.

Prompt engineering is one way to improve LLM performance~\cite{chen2023unleashing}. One such technique is chain-of-thought (CoT) prompting, shown to be effective for reasoning tasks~\cite{wei2022chain}. CoT prompting uses intermediate reasoning steps to break down the problem at hand, which provides more information on the process that was used to arrive at a given answer.

For code translation specifically, improvements in code representation through natural language (NL) and abstract syntax trees (ASTs) have been reported for software engineering tasks~\cite{neamtiu2005understanding,tang2021ast}. For example, summarizing code to NL leads to better code translation performance~\cite{ahmad-etal-2023-summarize}. Similarly, employing AST in code summarization outperforms state-of-the-art models~\cite{tang2021ast}.

Motivated by these observations, we investigate whether LLM-driven code translation can benefit from NL and AST intermediate representations (IRs) when combined with suitable prompt engineering. While there has been work on leveraging IRs for translation~\cite{ahmad-etal-2023-summarize,huang-etal-2023-program} and on LLM-based translation~\cite{pan2024lost}, our work is the first to systematically combine LLMs with both forms of IRs for code translation.

We explore two approaches to incorporate IRs in code translation:
(1)~a two-step approach, where the LLM first translates the original code to IR and then translates this IR to the target language~\cite{ahmad-etal-2023-summarize}; and 
(2)~a CoT prompting approach, where the LLM is instructed to use IR to explain its reasoning during translation.
We experiment with these two approaches using different permutations of IRs (NL, AST, or both), and compare with the simple zero-shot and one-shot prompt baselines.
Experiments using Open GPT4 8X7B and specialized StarCoder and CodeGen models on code translation benchmarks (CodeNet and AVATAR) show that CoT with an intermediate NL representation produces the most successful translations, with an increase of 13.8\% and 6.7\% on CodeNet and AVATAR, respectively, compared to the zero-shot prompt.

\vspace{3pt}
\noindent
Our main contributions in this work include:
\begin{itemize}
\item The first systematic study combining LLMs with natural language (NL) and abstract syntax trees (AST) as intermediate representations (IRs) for code translation.
\item Evaluation across multiple LLMs (Open GPT4 8X7B, StarCoder, CodeGen), datasets (CodeNet, AVATAR), and prompting strategies (zero-shot, one-shot, two-step, and CoT).
\item Our results show that CoT with NL summaries achieves the best performance, improving code translation accuracy by up to 13.8\% compared to the zero-shot baseline.
\end{itemize}

\noindent
Our code is available at:\\
\url{https://github.com/catai9/nl-in-middle/}

\begin{figure*}[!h]
    \centering
    \includegraphics[width=\linewidth]{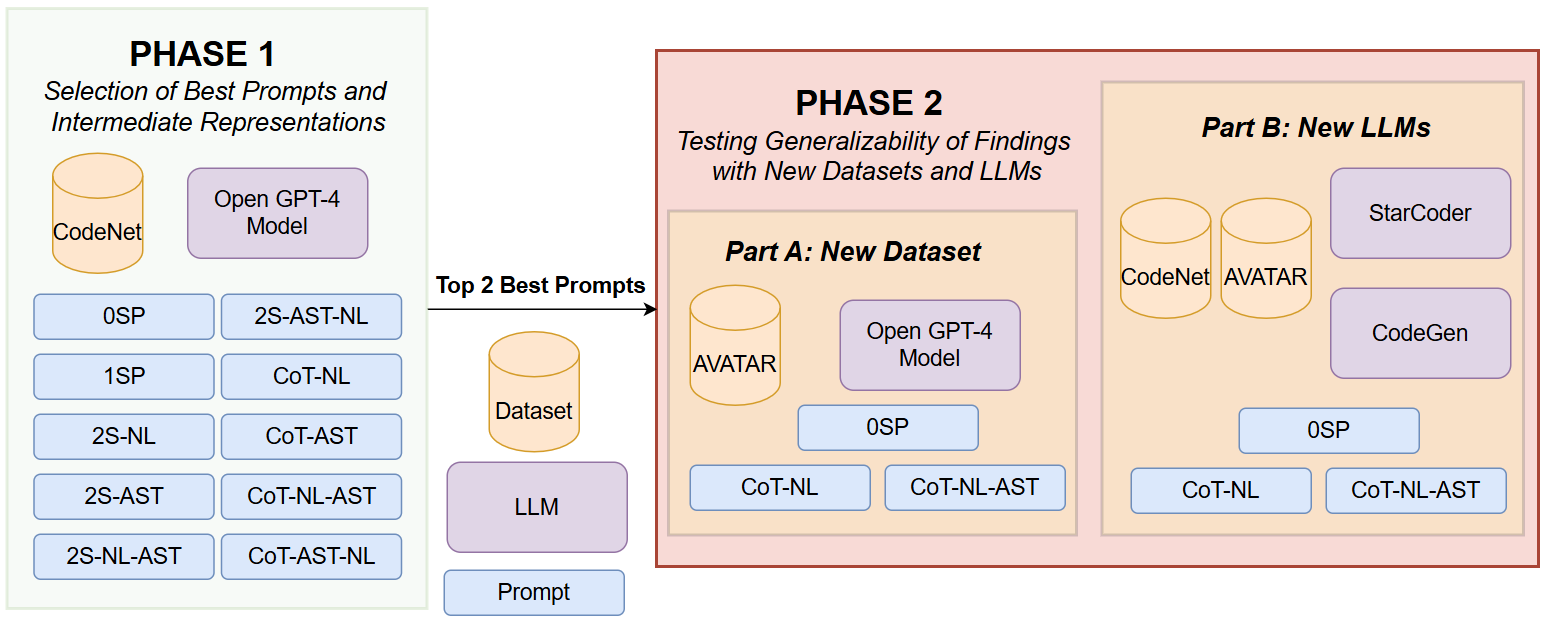}
    \caption{Overview of our methodology, with Phase 1 exploring ten prompt types and Phase 2 leveraging the two best prompts from Phase 1 to explore the generalization on a new dataset and with new LLMs.}
  \label{fig:process-map-overview}
\end{figure*}

\section{Related Work}
We categorize related work into code translation with LLMs and code translation with intermediate steps. Combining these two directions, i.e., code translation with LLMs and intermediate representations, is the novelty of our work.

\subsection{Code Translation with LLMs}
\label{related_work:llm}

Pan et al.~\cite{pan2024lost} conducted a comprehensive study of LLM-based code translation across five languages (C, C++, Go, Java, and Python). They evaluated multiple models including specialized code LLMs (CodeGen~\cite{nijkamp2023codegenopenlargelanguage}, StarCoder~\cite{li2023starcodersourceyou}, and CodeGeeX~\cite{zheng2024codegeexpretrainedmodelcode}), general-purpose LLMs, and proprietary GPT-4~\cite{openai2024gpt4technicalreport}. The authors also defined a metric to capture the percentage of successful translations, requiring the translated code to compile, pass runtime checks, and pass existing tests. They reported successful translation rates ranging from 2.1\% to 47.3\% across studied LLMs, with GPT-4 achieving the best performance.
The authors also adjusted the LLM prompt in the case of unsuccessful translation by providing the previous translation in the prompt, stating that it was incorrect, and asking for a re-generation, which led to more successful translations.  However, the use of intermediate representations was not considered.

Prior work has explored two main prompting strategies for LLM-based code translation. 
Few-shot learning improves model performance by providing custom examples to guide output style~\cite{pmlr-v70-finn17a,NIPS2017_cb8da676}; beyond code translation, it has also been applied to other software engineering tasks such as program refactoring and code review~\cite{10479398,pornprasit2024fine,LI2024112002}. 
Chain-of-thought (CoT) prompting, originally developed to be effective for math word problems~\cite{wei2022chain}, has been adapted for software engineering tasks including vulnerability discovery~\cite{nong2024chainofthoughtpromptinglargelanguage} and software architecture recovery~\cite{10.1145/3639476.3639776}. 
However, manually creating effective CoT prompts for tasks such as code generation can be challenging~\cite{10.1145/3690635,10634302}, which led to the development of three potential adaptions. The first approach, referred to as Structured CoTs, leverages program structures such as loops and branches to systematically create CoTs. This method organizes the reasoning process in a way that mirrors common programming practices, helping to structure and guide the thought process in a coherent manner~\cite{10.1145/3690635}. The second, COTTON~\cite{10634302}, uses lightweight language models for automatic CoT generation. The third, Tree of Thoughts (ToT)~\cite{yao2024tree}, explores multiple reasoning paths, but it tends to be slower and less effective for solving GitHub issues~\cite{larosa2024githubissuessolvedtree}. None of these extensions leverage intermediate representations, which is the focus of this work.

\subsection{Code Translation with Intermediate Steps}
Prior work has explored several kinds of intermediate representations for code translation. 
Ahmad et al.~\cite{ahmad-etal-2023-summarize} added a natural language summary to a sequence-to-sequence model, where code is first summarized and then the summary is used to generate code in the target language. This approach achieved competitive results on GitHub and CodeNet datasets. 
Szafraniec et al.~\cite{szafraniec2022code} used a compiler intermediate representation to improve translation performance. 
Huang et al.~\cite{huang-etal-2023-program} used abstract syntax trees (ASTs) as intermediate code representations. 
ASTs have also proven to be a valuable intermediate step in other software engineering tasks, such as code evolution, code summarization, and cross-language program classification~\cite{10.1145/1083142.1083143,zügner2021languageagnosticrepresentationlearningsource,9678882,8812062,10.1145/3524610.3527915}. This utility was further demonstrated in experiments using the CodeXGLUE and TransCoder translation benchmarks, where Huang et al.~\cite{huang-etal-2023-program} reported an average absolute improvement of 12.7\%.

\section{Methodology}
\begin{figure*}[t]
    \centering
    \includegraphics[width=\linewidth]{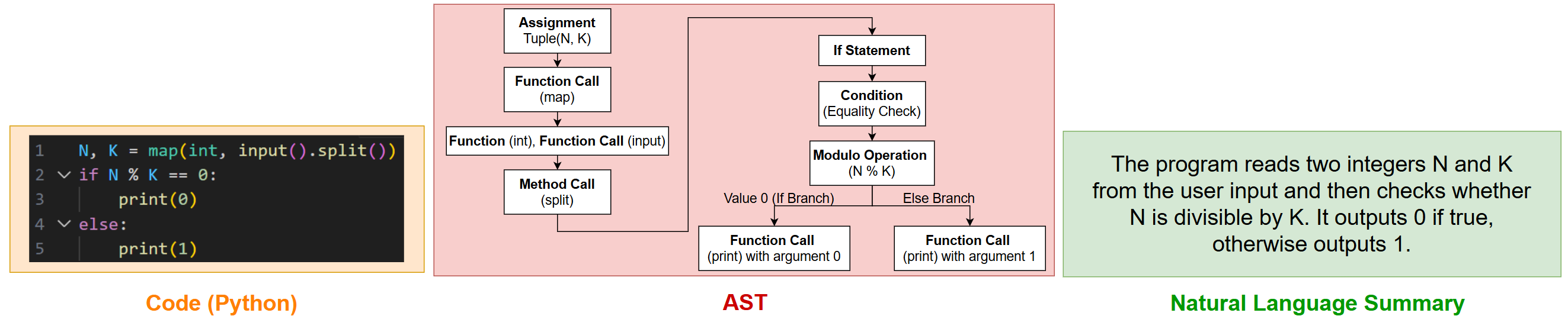}
    \caption{Sample code snippet with its AST and NL summary.}
  \label{fig:representations}
\end{figure*}

As illustrated in Figure~\ref{fig:process-map-overview}, we 
compare the performance of various prompts with and without intermediate code representations (Phase 1).  Following work by Pan et al.~\cite{pan2024lost}, we use the percentage of successful translations as the performance metric, where a translation is ``successful'' if the code compiles, passes runtime checks, and passes existing tests (recall Section~\ref{related_work:llm}).  We then test the generalizability of the two best prompts with an additional dataset (Phase 2A) and additional models (Phase 2B). Three NVIDIA RTX 6000 GPUs were used,  with 51.5 GB memory each. All prompts were evaluated with a temperature of 0.2 to reduce randomness of the outputs. Postprocessing was also conducted on the LLM outputs to remove the initial first line if it starts with ``Here is'' or ``Here's'', trim the first line of the file to remove any whitespace prefix, and delete any line that started with ``End of Code''.

In both Phase 1 and 2, to enable comparisons with prior work, we use the same dataset as in Pan et al.~\cite{pan2024lost}, i.e., a sample of size 1,000 from the CodeNet dataset~\cite{puricodenet}, a popular code translation benchmark. This sample contains 200 code snippets each of C++, C, Python, Java, and Go.  We run the experiments across all combinations of these languages (i.e., translating the 200 C++ code snippets to C, C++ to Python, C++ to Java, C++ to Go, C to C++, C to Python, etc.), resulting in 4,000 combinations. 

The CodeNet dataset was created by consolidating submissions from two programming contest web sites (AIZU Online Judge~\cite{aizu} and AtCoder~\cite{atcoder}) and contains code from over 50 programming languages. 

\subsection{Phase 1: Selection of Best Prompts and Intermediate Representations}
In Phase 1, we examine LLM-driven code translation through the use of various intermediate representations. Figure~\ref{fig:representations} shows a piece of code and its corresponding AST and NL representations. The AST shows the structure of the code snippet in a tree-style format and is derived by using the `ast' method in Python. Notably, the AST must be flattened (converted to text) for input to an LLM. The NL summary describes the code in plain text. 

In this phase, we evaluate the performance of the following options using the HuggingFace version of the open-source GPT-4 as the backbone (Open GPT4 8X7B~\cite{theblokeopengpt4}). The specific prompts that were used can also be found at \url{https://github.com/catai9/nl-in-middle/}.

\begin{itemize}

\item Zero-shot prompt (0SP), corresponding to previous work on code translation \cite{pan2024lost}.

\item One-shot prompt (1SP), with one example of source-to-target language translation for each source-target pair, randomly selected from the CodeTransOcean dataset~\cite{yan-etal-2023-codetransocean}.

\item A zero-shot two-step approach of having the LLM translate the source language to either AST (2S-AST), NL (2S-NL), both AST and NL (2S-AST-NL), or both NL and AST (2S-NL-AST), followed by having the LLM translate the intermediate representation to the target language.  

\item A one-shot CoT prompt in which the LLM is instructed to use either AST (CoT-AST), NL (CoT-NL), both AST and NL (CoT-AST-NL), or both NL and AST (CoT-NL-AST) to explain its reasoning. The AST was obtained using the Python `ast' method whereas the NL was obtained from the Rosetta Code website~\cite{rosettacode}. The same example used in the 1SP was also used here.
\end{itemize}

We test different combinations of AST and NL, given that prior work highlighted the importance of order in prompting~\cite{kumar2021reordering}. 

\subsection{Phase 2: Testing Generalizability with New Dataset and New LLMs}

\textbf{New Dataset.} After comparing the prompts using Open GPT4 8X7B on the CodeNet dataset in Phase 1, we gauge the generalizability of our findings with another dataset in Phase 2A. We use the AVATAR dataset here~\cite{ahmad2023avatarparallelcorpusjavapython} due to its large size and the availability of tests to allow for the computation of the percentage of successful translation metric. AVATAR contains 249 Java code snippets and 250 Python code snippets. We experiment with translating these snippets to C++, C+, Go, Python (only for Java snippets), and Java (only for Python snippets) resulting in 1,996 total combinations. Other datasets such as CodeTransOcean~\cite{yan-etal-2023-codetransocean} and CodeTrans from CodeXGLUE~\cite{lu2021codexgluemachinelearningbenchmark} do not include tests that could be used for judging whether the translation was successful and instead rely on metrics such as exact match, BLEU~\cite{papineni2002bleu}, and CodeBLEU~\cite{ren2020codebleu}, techniques that compare token n-grams in the generated text with the original reference.  These metrics may be inappropriate for code translation: high n-gram similarity does not necessarily imply code correctness in the presence of subtle bugs and low n-gram similarity might be due to a different way to write a correct program \cite{pan2024lost}. 

\textbf{New Models.} Finally, in Phase 2B, we investigate how the two best prompts from Phase 1 compare against the baseline prompt (zero-shot) for specialized open-source LLMs. We run these three prompts (zero-shot, top two best prompts from Open GPT4 8X7B on CodeNet) for these LLMs (StarCoder, and CodeGen) on both the CodeNet~\cite{puricodenet} and AVATAR ~\cite{ahmad2023avatarparallelcorpusjavapython} datasets to investigate the transferability of the outcomes. 

\section{Results}
\subsection{Phase 1}
Table~\ref{tab:postprocessed-results} shows that our results for Open GPT4 8X7B, averaged across all source-target language pairs, differ from those reported by~\cite{pan2024lost}, which reported an average of 82.0\% for the CodeNet samples for the same zero-shot prompt with the proprietary GPT-4 model. This difference could be attributed to the enhancements that are routinely added to GPT-4 and the general variability of LLM results. Notably, the zero-shot prompt results for Open GPT4 8X7B are still higher than those reported by the majority of the evaluated LLMs in~\cite{pan2024lost}. Furthermore, the best prompt result from Open GPT4 8X7B (CoT-NL) outperforms all of the open-source reported models in~\cite{pan2024lost}. 

\begin{table}[t]
    \centering
    \small
    \caption{Performance of Open GPT4 8X7B on CodeNet compared to the results obtained by Pan et al.~\cite{pan2024lost}, with the two prompts achieving the best Open GPT4 8X7B results bolded; Success\% = successful translation rate.}
    \label{tab:postprocessed-results}
    \begin{tabular}{lr}
    \hline
    \multicolumn{1}{c}{\multirow{1}{*}{\textbf{Prompt}}} & \multicolumn{1}{c}{\textbf{Success\%}} \\
    \hline
    \textit{Open GPT4 8X7B} & \\
    \hline
    0SP & 28.6\% \\
    1SP & 33.5\% \\
    2S-NL & 10.7\% \\
    2S-AST & 2.6\% \\
    \textbf{CoT-NL} & \textbf{42.4\%} \\
    CoT-AST & 32.2\% \\ 
    2S-NL-AST & 9.0\% \\
    2S-AST-NL & 11.2\% \\
    \textbf{CoT-NL-AST} & \textbf{39.6\%} \\
    CoT-AST-NL & 37.9\% \\
    \hline
    \textit{Other LLMs from Pan et al.~\cite{pan2024lost}} & \\
    \hline
    CodeGen & 18.1\% \\
    CodeGeeX & 8.4\% \\ 
    StarCoder & 37.3\% \\
    Proprietary GPT-4 & 82.0\% \\
    Llama 2 & 13.2\% \\ 
    TB-Airoboros & 9.3\% \\
    TB-Vicuna & 2.0\% \\
    \hline
    \end{tabular}
\end{table}

CoT-NL generally achieves the best performance, followed by CoT-NL-AST. Similar to prior works on LLM few-shot learning~\cite{kumar2021reordering}, we also found that order matters: NL-AST achieved better results than the order AST-NL in the prompt. 

\subsection{Phase 2A}
Using the baseline zero-shot prompt and the top two prompts from Phase 1 (CoT-NL and CoT-NL-AST), we now compare performance on the AVATAR dataset~\cite{ahmad2023avatarparallelcorpusjavapython} to gauge the generalizability of our findings. Table~\ref{tab:llm-results} shows that Open GPT4 8X7B also has the best performance using the CoT-NL prompt for the AVATAR dataset
and CoT-NL-AST outperforms the zero-shot prompt by 4.6\%. 

\begin{table}[t]
    \centering
    \small
    \caption{Comparison of the top two best prompts against the zero-shot prompt for both the CodeNet and Avatar datasets for the three studied models, with the best result highlighted in blue}
    \label{tab:llm-results}
    \begin{tabular}{l l c c}         
    \hline
    \multirow{2}{*}{\textbf{Model}} & \multirow{2}{*}{\textbf{Prompt}} & \multicolumn{2}{c}{\textbf{Success\%}} \\
     &  & \textbf{CodeNet} & \textbf{AVATAR} \\
    \hline
    \multirow{3}{*}{Open GPT 4 8X7B} & 0SP & 28.6\% & 17.6\% \\
     & CoT-NL & \textcolor{blue}{\textbf{42.4\%}} & \textcolor{blue}{\textbf{24.3\%}} \\
     & CoT-NL-AST & 39.6\% & 22.2\% \\
     \hline
     \multirow{4}{*}{CodeGen} & Pan et al.~\cite{pan2024lost} & 18.1\% & 5.9\% \\
     & 0SP & 18.4\% & 6.8\% \\
     & CoT-NL & 4.9\% & 1.3\% \\
     & CoT-NL-AST & 2.9\% & 0.2\% \\
     \hline
     \multirow{4}{*}{StarCoder} & Pan et al.~\cite{pan2024lost} & 37.3\% & 13.1\% \\
     & 0SP & 36.3\% & 20.4\% \\
     & CoT-NL & 38.0\% & 20.6\% \\
     & CoT-NL-AST & 31.2\% & 16.9\% \\
     \hline
    \end{tabular}    
\end{table}
    
\subsection{Phase 2B}
Experiments with other models -- CodeGen and StarCoder -- for both the CodeNet and AVATAR datasets show differing conclusions. First, we run the baseline zero-shot prompt from~\cite{pan2024lost} and achieve similar performance on the zero-shot prompt for both CodeGen and StarCoder on the CodeNet and AVATAR datasets. This is shown in Table~\ref{tab:llm-results}. Although the zero-shot performance with StarCoder on the AVATAR dataset was 7.3\% higher than that reported by~\cite{pan2024lost}, the improvement exists for both Java and Python source languages and can be attributed to the inherent randomness in LLM outputs.

For the CodeGen model, Table~\ref{tab:llm-results} shows that the zero-shot prompt performs the best compared to CoT-NL and CoT-NL-AST. For the StarCoder model, CoT-NL generally achieves better percentages than the zero-shot prompt, but CoT-NL-AST performs worse than the zero-shot. However, Open GPT4 8X7B still obtains the highest percentage of successful translations compared to both CodeGen and StarCoder.

Open GPT4 8X7B has the fewest reported parameters of 7 billion, with a model size of 26.44 GB~\cite{theblokeopengpt4}. On the other hand, StarCoder has 15.5 billion parameters and a model size of 59.19 GB~\cite{li2023starcodersourceyou} whilst CodeGen has 16.0 billion parameters and a model size of 61.16 GB~\cite{nijkamp2023codegenopenlargelanguage}. Although StarCoder initially obtained a higher percentage of successful translations for the zero-shot prompt compared to the Open GPT4 8X7B model, Table~\ref{tab:llm-results} shows that an enhanced prompt (CoT-NL) can result in better performance with a smaller model.

\section{Discussion}
\subsection{Limitations}
One limitation of our work is the focus on open-source LLMs. According to Pan et al.~\cite{pan2024lost}, the proprietary GPT-4 model achieved significantly better results compared to the other approaches. However, we find our work to be more reflective of enterprise use cases for code translation as many companies have policies against using proprietary LLMs with sensitive data (such as company code)~\cite{forbesAppleJoins,hackernoonCompaniesBanning,linkedinEmployersMust}. Our study shows that leveraging an open-source LLM could be beneficial for code translation and these open-source models could also operate locally, assuming the hardware requirements are met. 

Another limitation may be due to our selection of datasets. Although we attempted to generalize the applicability of our findings by investigating two open-source datasets, it is possible that these results are invalid on other datasets. On the other hand, these two datasets were the only ones with test cases. Similarly, we only experimented with three open-source LLMs, but other open-source LLMs could have better performance or showcase a different pattern for the prompt formats. That said, the selection of these open-source LLMs was based on prior research comparing the different open-source LLMs, so it is likely that these three are the best ones available. 

Furthermore, our work is limited to the prompts that we tested. Although we attempted to create a representative prompt for each investigated technique, it is possible that better prompts may exist for each category. To mitigate this limitation, we also experimented with other formats and improvements such as using whitespace and custom examples for improving our prompts; however, we did not observe any performance improvements.  Finally, we also focus on only five programming languages and it is possible that the behavior would differ across other programming languages. 

\subsection{Future Work}
In this paper, we assess the quality of the code translation using a binary metric that measures whether the translation was successful, but future work should explore more robust evaluation metrics that penalize more for significant functional errors rather than subtle syntactic errors (e.g., missing imports) that disproportionately affect the binary metric. 

Future studies can also expand this work by considering additional languages beyond C++, C, Python, Java, and Go, including older languages such as COBOL, Pascal, and Fortran for legacy code translation. Alternatively, the study can be extended to focus specifically on test generation performance, particularly in creating effective unit and integration tests for production-level code. With the rising popularity of agents, future work can also examine using agentic artificial intelligence to improve code generation quality where the agent automatically corrects syntactic and minor errors that do not affect functional correctness.

Our work demonstrates that the general-purpose GPT-4 8x7B LLM outperformed larger, code-specialized LLMs; however, open-source models still lag behind the reported performance of proprietary models. Future research should investigate the reasons behind the performance gap between general-purpose and code-specialized LLMs, as well as between proprietary and open-source models, to better understand key architectural or training differences.

In addition, experimentation with few-shot learning and variations of CoT prompting (e.g., Structured CoTs~\cite{10.1145/3690635}, COTTON~\cite{10634302}) could also be conducted to determine if intermediate representations would also aid in increasing performance for code translation. 

Other future work could be to improve the post-processing of the LLM output along with trying more specific intermediate code representations during code translation. In this work, we post-processed the LLM output to reduce the frequency of noisy non-code lines. However, we only addressed the most common issues using a simple heuristic check, but there could be more specific post-processing to boost the percentage of successful translations. Another lens of future work would be to leverage more specific intermediate code representations. In our prompt formats, we focused on NL and a general AST but there has been progress on self-optimizing ASTs to incorporate specific language nuances~\cite{wurthinger2012self}. Control flow graphs~\cite{1702388} are another option that could be used as an intermediate code representation.

\section{Conclusions}
We studied the effects of introducing an intermediate code representation step in code translation with LLMs. The motivation of this work stemmed from improvements in prompting techniques such as CoT and enhanced code representation frameworks such as ASTs. We experimented with different prompts such as two-step prompts and CoT prompts with natural language and AST intermediate code representations, and compared the performance using Open GPT4 8X7B, StarCoder, and CodeGen models on the CodeNet and AVATAR datasets. We found that CoT with natural language representation performed the best, with a 13.8\% and 6.7\% improvement on the CodeNet dataset and AVATAR dataset, respectively, compared to the initial zero-shot prompt with Open GPT4 8X7B. Our experiments with various models and datasets demonstrate the potential for generalizing our findings while also emphasizing the effectiveness of incorporating an intermediate code representation in code translation.

\bibliographystyle{IEEEtran}
\bibliography{custom}

\end{document}